\documentclass[]{article}

\usepackage{color}
\usepackage[usenames,dvipsnames,svgnames,table]{xcolor}
\usepackage{graphicx}
\usepackage{subfig}

\small

\newcommand{\Msun} {\mbox{$~M_{\odot}$}}

\begin{document}
\sloppy

\twocolumn

\begin{center}
\fboxrule0.02cm
\fboxsep0.4cm
\fcolorbox{Brown}{Ivory}{\rule[-0.9cm]{0.0cm}{1.8cm}{\parbox{7.8cm}
{ \begin{center}
{\Large\em Perspective}

\vspace{0.5cm}

{\Large\bf Planet formation in clusters}

\vspace{0.2cm}

{\large\em Susanne Pfalzner}


\vspace{0.5cm}

\centering
\includegraphics[width=0.23\textwidth]{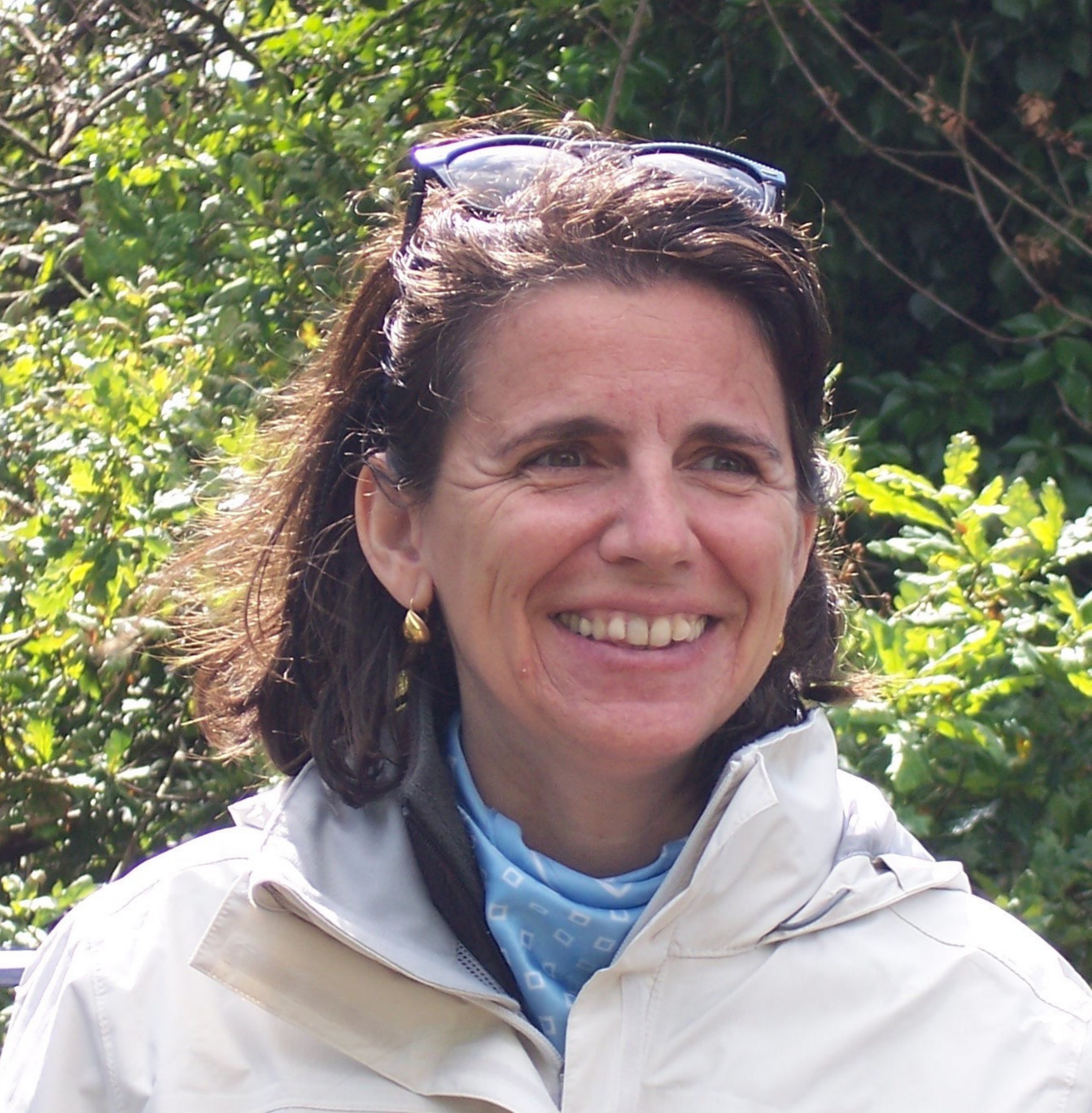}
\end{center}
}}}
\end{center}

One well-tested method in science is to separate the object of interest from its surroundings and look at it in isolation. The advantage is that unimportant  information is removed and the true properties of the object are seen more clearly.
However, sometimes the influences of the surroundings actually determine the properties of an object. In this case, not taking the environment into account can lead to incomplete or even false conclusions.

In the context of planet formation this question arises to: is it sufficient to study the nascent planetary system in isolation?  Stars usually do \emph{not} form in isolation but as part of a stellar group (Lada \& Lada 2003, Porras 2003). The first important question in this field is then:~\emph{How important is the influence of the surrounding stars on circumstellar discs and forming planetary systems?} Distance  to other stars in the same system is the key factor here, so the main parameter is obviously the local stellar density.

There are basically two main channels of influence exerted by the surrounding stars on discs: their radiation and gravitational interactions. Both effects might truncate or possibly even completely destroy a planet-forming disc. Famous examples of the influence of radiation on discs are the so-called proplyds in Orion, which show the destructive power of external photo-evaporation on young discs (Fig.~1a). The signs of gravitational truncation of discs are more difficult to track because close flybys mainly occur in the strongly embedded, and therefore difficult to observe, phase of young clusters.  Tell-tale signs are spiral arms (Fig.~1b). However, they can also be caused by other processes and tend to evolve into ring-like structures within just a few thousand years (Cuello et al.~2019). Thus, we can pose a second important question:~\emph{Does radiation or gravitational interaction dominate?}

\begin{figure}[h]
\includegraphics[width=0.49\textwidth]{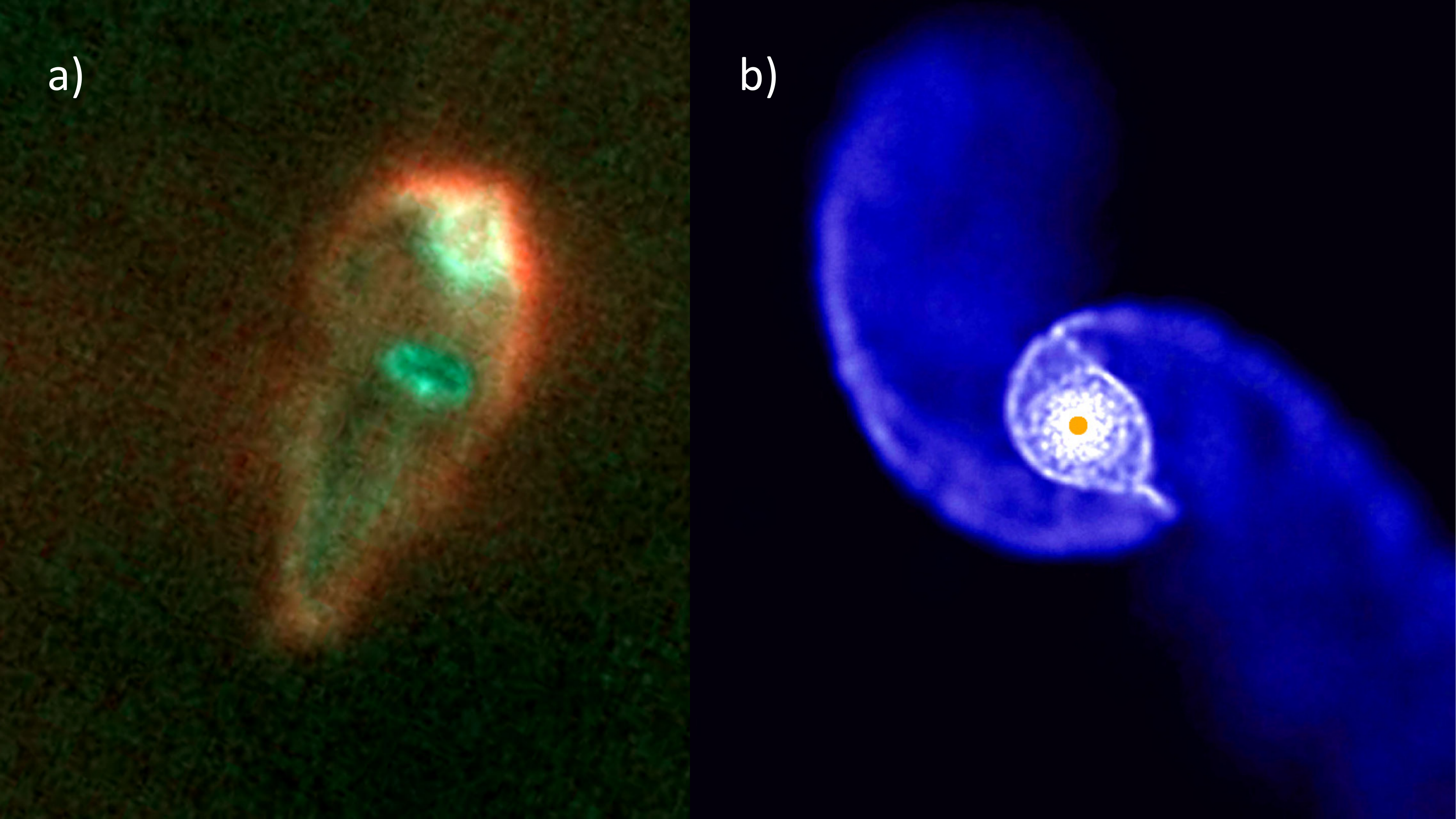}
\caption{Environmental effects on protoplanetary discs.  a) HST image of an externally photo-evaporated disc (credit: NASA/ESA) and b) simulation result of the effect of a stellar flyby.}
\label{fig:fig1}
\end{figure}

For both questions still no definite answer exists, but there has been considerable progress during past years. 

The central parameter is the ``stellar density" since it determines both the frequency and the strength of the external influence. This applies equally to radiation as to gravitational interactions. Sparse stellar groups like Cha I consist of just a few dozens of stars, whereas dense groups like NGC 3603  and Trumpler 14 can contain tens of thousands of stars packed within less than 1 pc$^3$. Even in high-mass stellar groups ($ M_c>10^3$\Msun) the average stellar densities vary over many orders of magnitude from \mbox{ $<$ 0.1 \Msun pc$^{-3}$} to $> 10^5$\Msun pc$^{-3}$ (Wolff et al.~2007).  

Even without large simulations it is intuitively obvious that environmental effects will play a much larger role in Trumpler 14 than in Cha I. This is why investigations to date have tended to concentrate on fairly dense systems like the Orion Nebula Cluster (ONC) due to its  relatively proximity to us (383 $\pm$ 3 pc; Kounkel et al.~2017). For this reason the ONC is often taken as model cluster for theoretical investigations, too (for example, Adams 2010, Wijnen et al.~2017, Winter et al.~2018, Cuello et al.~2019, Portegies Zwart 2019). The general outcome of these investigations is that the environment is only responsible for the destruction of 5-10\% of all discs. Some small disc sizes can be attributed to the environment, but typically the disc size in the ONC is only reduced to about 100 AU. This is only slightly smaller than the typical disc sizes of 100-200 AU observed in sparse associations.  

However, the density in a stellar group is far from constant, and changes by several orders of magnitude within just a few Myr. In my opinion, the key here lies in understanding early cluster dynamics. This means that the result obtained for the ONC is far from universal: it is only valid for this particular stellar group at this specific phase of its temporal development. If one is interested in the environmental influence on discs in general, one has to look at different stellar groups and also consider the time  dependence of the stellar density.  

Fortunately, we are spared the mammoth task of considering every observed stellar group individually, because it turns out there seem to be only two types of stellar groups: those that develop into long-lived clusters and those that largely dissolve within about 5-10 Myr. These two types are often referred to as clusters and associations, respectively\footnote{Historically it was thought that clusters formed as bound and associations as unbound systems.  Nowadays we know that most stellar groups are bound as long as they are still embedded in gas. The historical misinterpretation leads to the confusing nomenclature that many embedded stellar groups are labelled clusters, even though they will largely dissolve as soon as they lose their gas like most of the stellar groups in the solar neighbourhood.}. Clusters are much more efficient in converting gas and dust into stars which allows them to form much more compact entities: at any given age their stellar density is at least 100$\times$ higher than that of an association of the same age and mass. However, it should be stressed that about 80-90\% of the field population was formed in short-lived associations and only 10-20\% in long-lived clusters (Adamo et al. 2011, Bastian 2013).

\begin{figure}[t]
\includegraphics[width=0.50\textwidth]{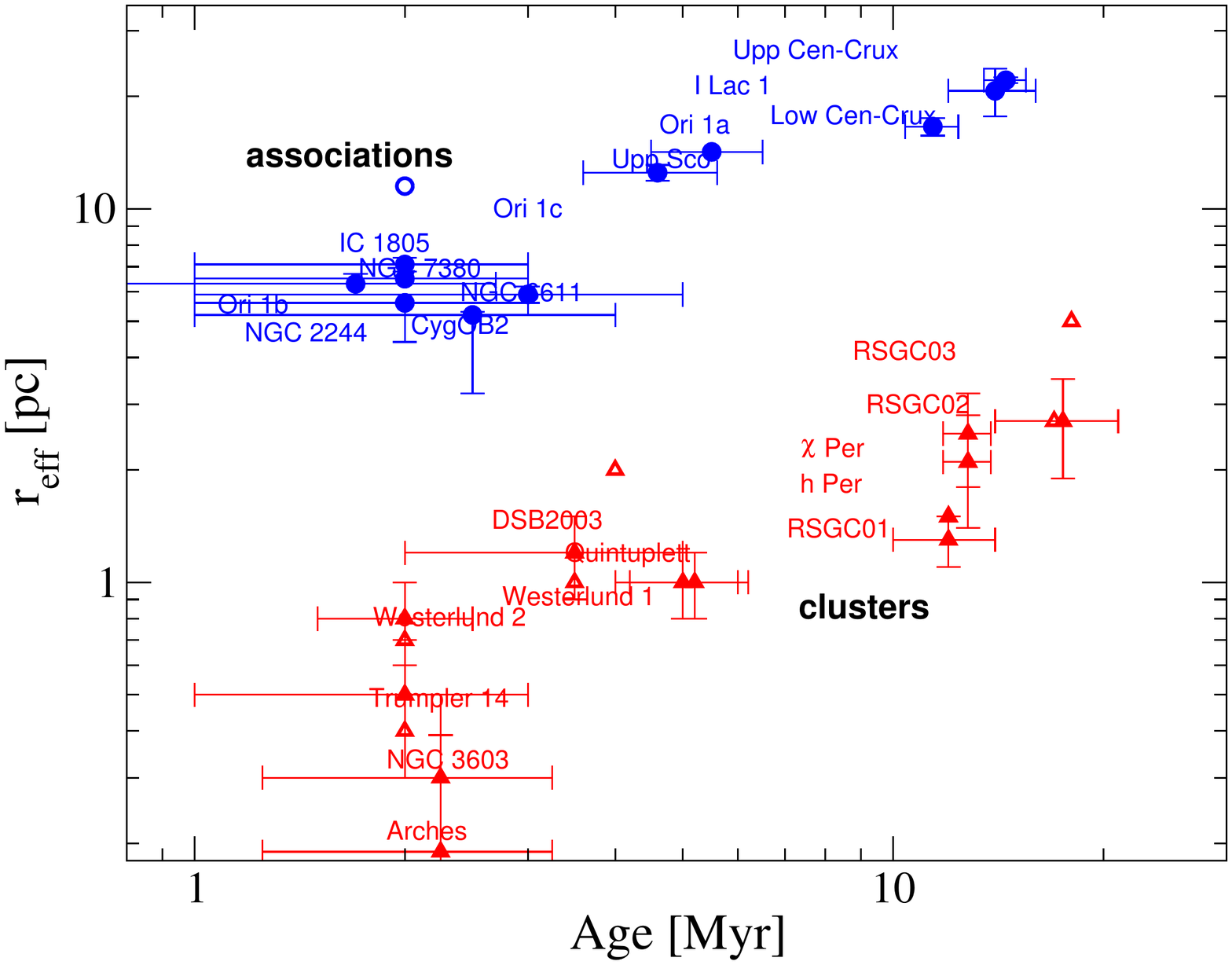}
\caption{Two types of stellar groups --- compact clusters (red) and loose associations (blue, figure adapted from Pfalzner 2009).} 
\label{fig:fig2}
\end{figure}

Associations and clusters develop along different, but well-defined evolutionary trajectories in terms of cluster size as a function of time (see Fig.~2). This means that simulations conforming to either of these temporal paths are not only valid for one specific stellar group at a given point of time, but for an entire class of stellar groups over the full time period considered --- a huge reduction in the possible parameter space.

Again, observations are snapshots in time, yet young stellar groups are highly dynamical objects whose density is constantly changing (Concha-Ram{\'\i}rez et al.~2019). Hence, understanding the dynamics of young clusters is the key to answering the question of how strong the influence of the surroundings actually is. 

For a long time it was unclear whether stellar groups formed by merging of subgroups, or just appeared the way we observe them, and whether they expand as soon as they have formed. Recently the advent of Gaia results clarified this situation. At ages 1--3 Myr, stellar groups typically expand with a velocity of 0.5--1 pc/Myr (Kuhn et al.~2019) and
remain in the very dense phase 3 Myr at most, and probably more likely only 1-2 Myr. This means that the stellar groups with little-to-no gas have in fact been much denser in the past, when the influence of the environment has also been stronger. Our simulations (Vincke \& Pfalzner 2018) have shown that as soon as the gas is expelled and the stellar groups start to expand the influence of the environment drops considerably.

\begin{figure}[t]
\includegraphics[width=0.49\textwidth]{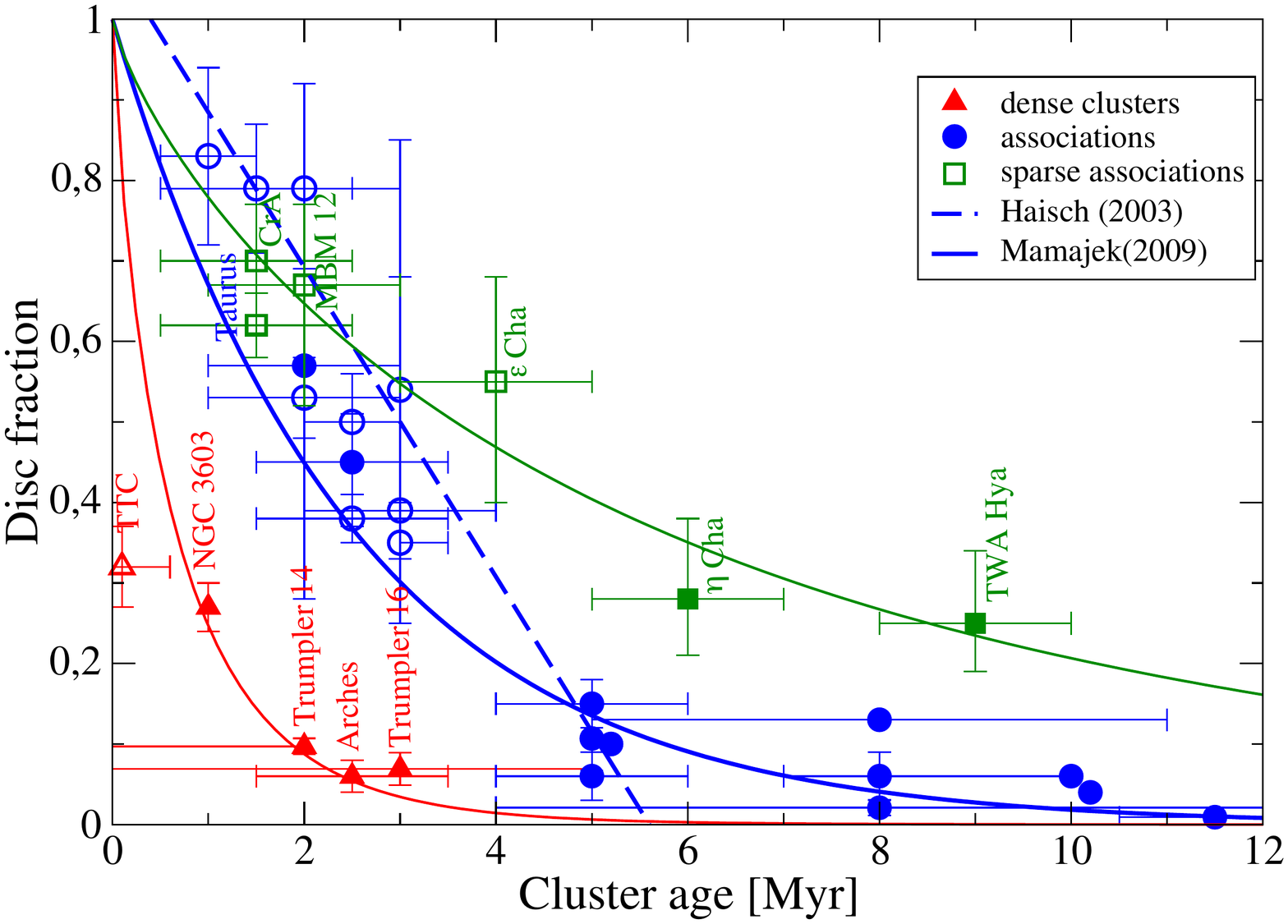}
\caption{Disc fraction in compact clusters (red), associations (blue) and co-moving groups (green), Pfalzner et al.~2014.}
\label{fig:fig1}
\end{figure}

Given the difference in density, environmental effects are generally much stronger in clusters than in associations. Two properties can be tested in this respect by simulations and observations alike: disc frequency and disc size. Roughly speaking it can be said that simulations show that complete disc destruction rarely happens in associations that contain only a few hundred stars; only in clusters that contain a few thousand or ten thousand stars does disc destruction become an issue, at least near the cluster centre. However, even in massive associations probably less than 20\% of discs can be completely destroyed by the environment (Olzcak et al.~2006, Vincke \& Pfalzner 2016). 

Conversely, in clusters complete disc destruction by the environment is a real factor.  A strong observational indicator is the much lower disc frequency in clusters compared to associations (see Fig.~3) at a given age. This seems to indicate that for very dense massive clusters environmental effects might be the dominant disc destruction process. However, one has to be careful with such a judgement as we have no observations covering the formation phase of such dense clusters. Therefore we do not know whether these clusters start out with basically all stars being surrounded by discs or not.  

Total disc destruction is of course an extreme and rather negative outcome as far as forming planetary systems are concerned! What is much more common is that the environment leads to disc truncation or redistribution of disc material. Just considering gravitational interactions, the high flyby frequency in stellar groups like the ONC could be the reason that discs are rarely larger than 100-200 AU in such environments (Vincke 2016).

\begin{figure}[t]
\includegraphics[width=0.49\textwidth]{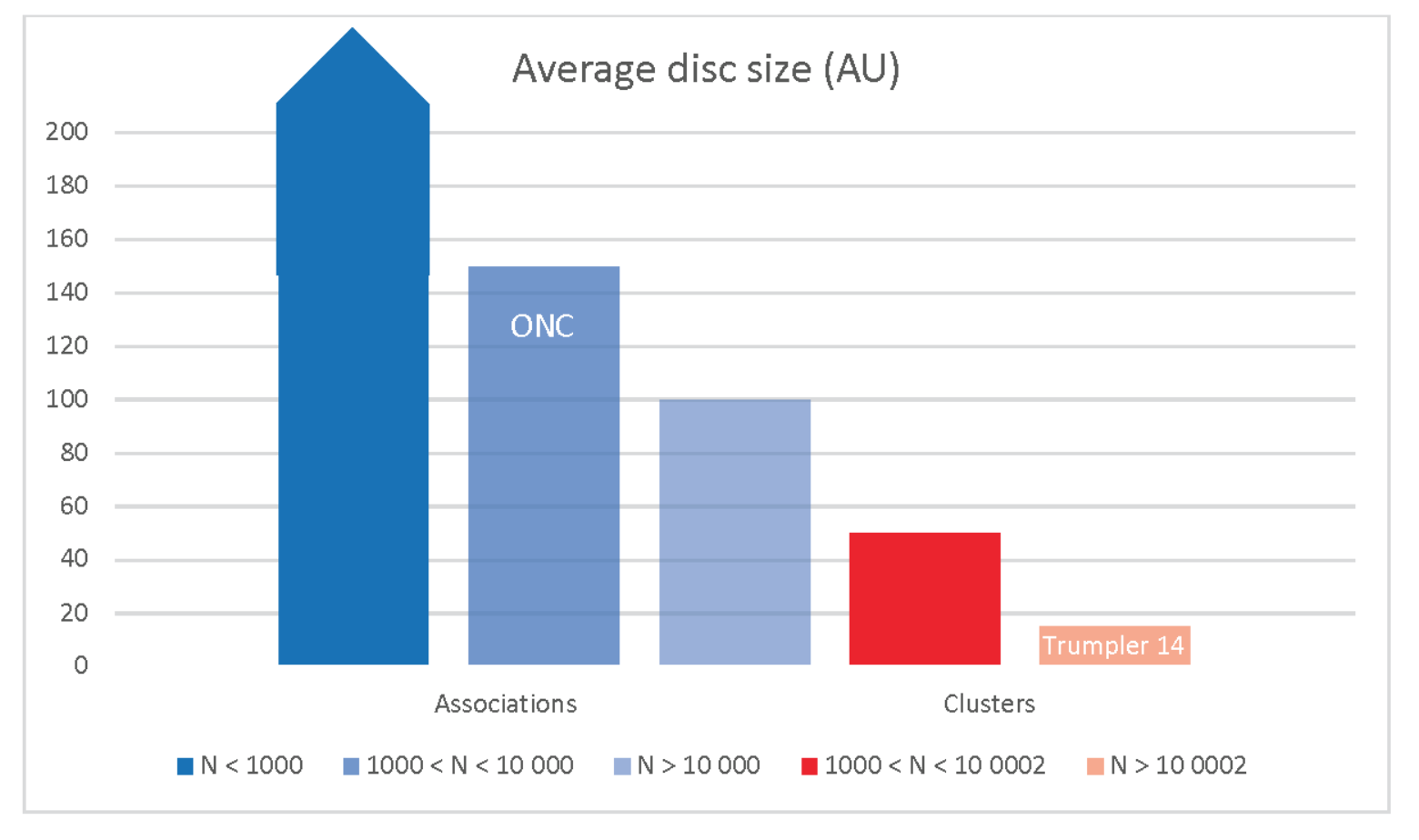}
\caption{Average disc sizes expected from simulations of different environments at ages 3 Myr and older. Summary of the results by Vincke \& Pfalzner 2016, 2018}
\label{fig:fig1}
\end{figure}

Unfortunately most clusters are further away from us than associations, so that there is little information about the disc sizes in such environments.  Fig.~4 gives a schematic illustration of the average disc sizes that are expected in different environments from simulations. In the typical stellar groups found in the solar neighbourhood containing just a few hundred stars, the disc size is probably hardly affected by the environment. By contrast, in dense long-lived clusters, the mean disc size should be of the order of just 10-20 AU in environments like Westerlund~2 and Trumpler~14. This means that planetary systems in clusters should be much more compact than those that formed around field stars (Fujii \& Hori 2019). In addition, it can be expected that the outermost planets of these planetary systems are often on eccentric and inclined orbits. As illustrated in Fig.~1b, any close stellar flybys excites matter onto eccentric orbits, which is not only the case for disc matter, but also for already formed planets. As most flybys will have some angle to the plane of the planets, this leads to inclined orbits. Pr 0211 in M44 is possibly a first example for such a system (Pfalzner et al.~2018).

The values given above are averages over entire stellar groups. However, it is the most central areas where the density is highest, and thus where disc destruction happens most frequently and disc fractions are lower than elsewhere. As these stellar groups expand so do these inner areas. As observations mainly focus on the similar-sized regions independent of cluster age, the cluster expansion might lead to a faster drop in disc fraction than really happens (Pfalzner et al.~2014) 

There has been a long debate about whether external photo-evaporation or gravitational interactions dominate in star forming regions (recently, Champion et al.~2017, Haworth et al.~2018, Winter et al.~2018). To me, the answer is just that ``it depends". Fig.~5 illustrates this in a qualitative way. Both effects require high stellar densities to be efficient. However, for external photo-evaporation to work it is also necessary for the radiation to penetrate efficiently through the cluster/association. This means that the gas/dust density in the stellar group has to be low enough not to absorb the radiation. The deeply embedded phase lasts less than 1 Myr and so far very few observational constraints exist.
However, during this early phase it is gravitational interactions that set the scene.
Afterwards the efficiency of close stellar flybys steadily decreases, because as soon as most discs are truncated to the disc size that is typical for this stellar group, events that truncate the disc even further become seldom. The frequency of events that change the disc size is reduced further when the stellar density eventually drops due to cluster expansion.  The situation might be different in massive long-lived clusters, but very few observational constraints exist so far.

\begin{figure}[t]
\includegraphics[width=0.49\textwidth]{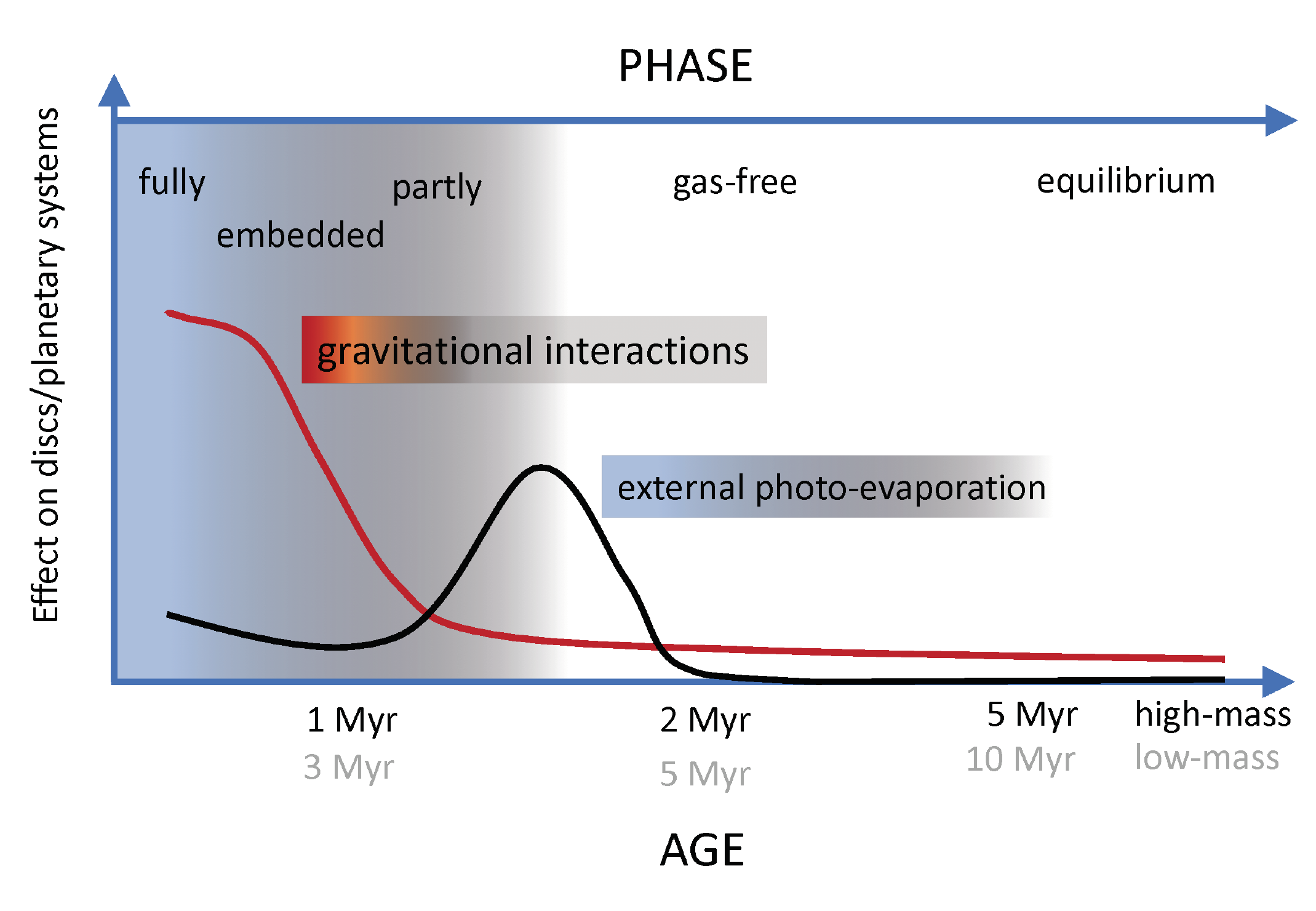}
\caption{Schematics of the relative importance of gravitational interactions versus external photo-evaporation during the different phases of cluster development.}
\label{fig:fig1}
\end{figure}

In associations external photo-evaporation (Fig.~5, black line) basically inherits the disc frequency and disc size distribution that close flybys have created and potentially leads to additional disc truncation or even destruction.  The proplyds in the ONC are a prime example of such a situation.  They are so clearly visible because the stellar group is just at the stage when it has become gas-free in the centre, while at the same time the stellar group has not yet started to expand.  This is the moment when external photo-evaporation is at its strongest.  However, this phase is relatively short-lived, because stellar groups are extremely dynamical. How short depends somewhat on the mass of the association: heavier ones develop faster than lighter ones.  Most of the environmental influence happens approximately during the first 2-3 Myr.  This means the time-window for efficient external photo-evaporation is only 1-2 Myr in associations. Afterwards, the massive stars dominating the radiation are just too distant to make external photo-evaporation work efficiently. 

By contrast, although the frequency of close flybys that actually have an effect on the discs decreases dramatically, they never completely vanish. This means even in the planetary system phase such interactions can also influence the newly formed planetary systems (Malmberg et al.~2007, Hao et al.~2013, Wang et al.~2015, Cai et al.~2017). This is especially true in the most massive clusters like for example Westerlund~1. For a long time it was thought that such dense environments would hinder the formation of planets or at least reduce it considerably. However, in recent years over twenty planets have been found in long-lived clusters (Pfalzner et al.~2018), so that at least for short-period planets the environment does not prevent their formation. However,  these systems will have likely more planets on highly eccentric orbits and fewer distant planets; something that future exoplanet surveys will test. 



The influence of the environment is not restricted to distant stellar groups: our own solar system was once shaped by its environment. Several properties like the sharp outer edge at 30 AU or the composition of meteorites (Adams et al.~2006, Adams 2010, Parker 2017, Portegies Zwart 2019) seem to bear the hallmarks of the past external influences. 
Even the orbits of the hundreds of thousands of small objects moving outside Neptune's orbit might be due to a close stellar flyby in the Sun's past (Pfalzner et al.~2018). 

The discovery of Oumuamua in 2017 (Meech et al.~2017) has brought a dramatic new aspect to the interplay between discs and their environment. Namely, what happens to all the material that becomes unbound due to the interactions with the stellar cluster (Hands et al. 2019)? If there are really 10$^{15}$ of such objects floating around in interstellar space, as anticipated, some of them could become part of a newly formed circumstellar disc. Due to their size they could help to jump the meter-sized barrier. This way it is possible that these small objects are vital agents in seeding or accelerating planetary growth (Pfalzner \& Bannister 2019). 

\begin{figure} 
\centering
\mbox{
\subfloat[]{\includegraphics[width=0.49\textwidth]{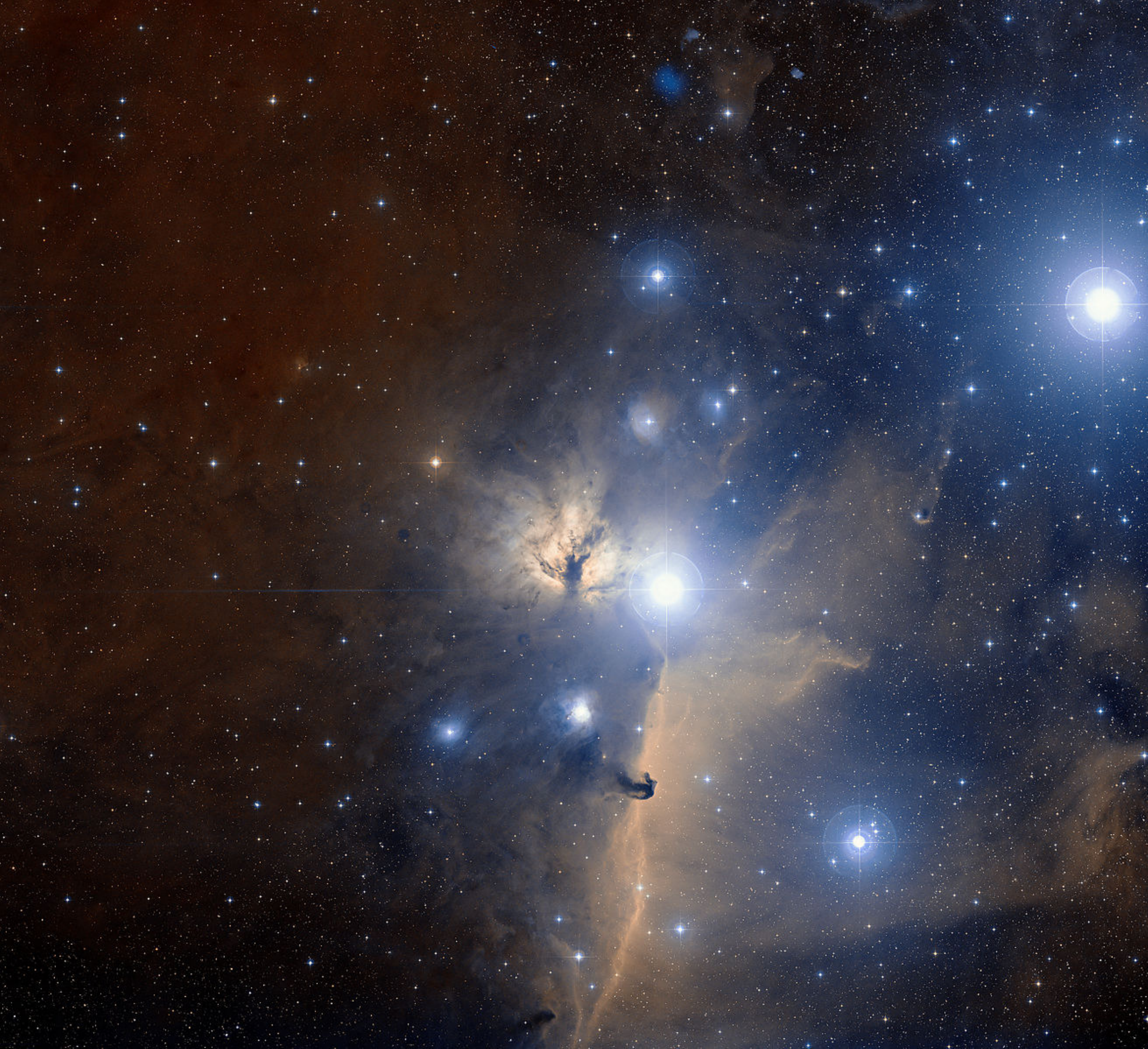}}}
\mbox{
\subfloat[]{\includegraphics[width=0.49\textwidth]{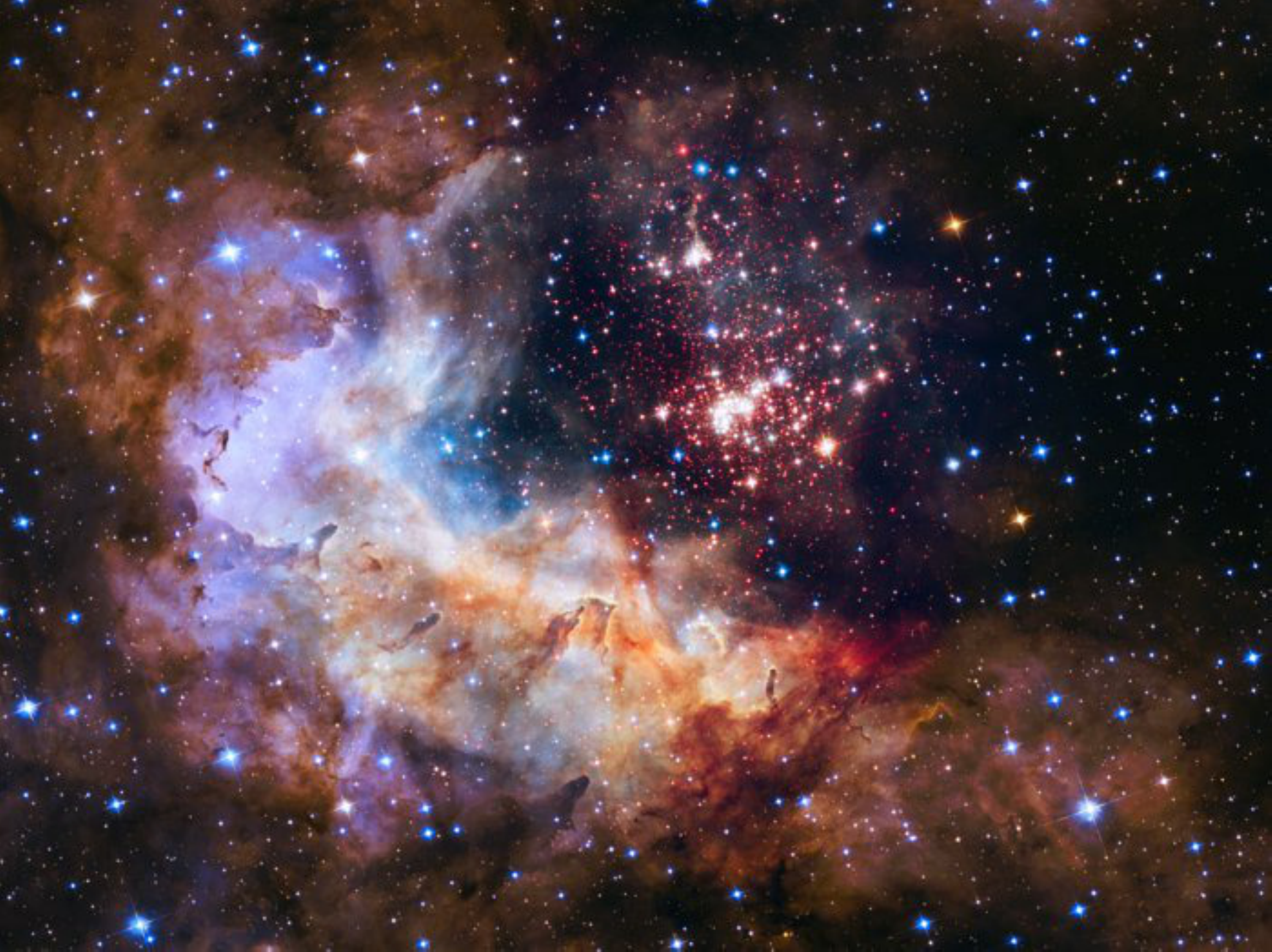}}}
\caption{Range of clustered star formation, stretching from (a) low-mass associations like $\sigma$ Ori, where the environment plays a minor role, to (b) high-mass clusters like Westerlund 2 where disc sizes are dominated by the environment. Image credits: ESO and Digitized Sky Survey 2; NASA, ESA, the Hubble Heritage Team (STScI/AURA), A. Nota (ESA/STScI), and the Westerlund 2 Science Team.}
\label{fig:last}
\end{figure}

In summary, I think it is important to see planetary systems not necessarily as isolated objects, but also to take the interactions with their environment into account. Star formation in the Milky Way mainly takes place in the short-lived associations, like the ones surrounding us. Here the influence on discs and forming planetary systems is non-negligible but moderate (Fig.~\ref{fig:last}a). By contrast, in long-lived clusters the environment plays a dominant role (Fig.~\ref{fig:last}b). Another important point is that even if specific clusters do not show signs of environmental influence going on at the moment, the situation might have been very different in their past. The relative importance of gravitational- vs. radiation-driven impact is a question of developmental phase. Gravitational interactions are most important during the strongly embedded phase, they set the initial conditions in terms of disc sizes and frequencies which can be modified by external photo-evaporation. However, there is only a relatively short time available for radiation to act.

What next? As pointed out above, cluster dynamics is a key factor here. The wealth of the Gaia data will hopefully enable us to put even tighter constraints on the dynamics of young clusters and associations alike. Equally, it would be important to obtain better information on discs in the embedded phase and in dense clusters, which is admittedly an observational challenge.  On the theoretical front, it is important to take the newly found observational constraints into account and eventually bridge the still existing gap between cluster formation models and the following phase of cluster expansion (Farias et al.~2019). Moreover, so far it is rarely taken into account that such clusters contain a large fraction of binaries and multiple systems, and future studies should include that. Given the considerable progress during the last few years, I think all these aims are reachable in the near future.

\bigskip

\normalsize

\footnotesize

{\bf References:}\\
Adamo, A., {\"O}stlin, G., \& Zackrisson, E.\ 2011, MNRAS, 417, 1904\\
Adams, F.~C., Proszkow, E.~M., et al.\ 2006, ApJ, 641, 504\\
Adams, F.~C.\ 2010, ARAA, 48, 47\\
Bastian, N.\ 2013, 370 Years of Astronomy in Utrecht, 287\\
Cai, M.~X., et al.\ 2017, MNRAS, 470, 4337\\
Champion, J., Bern{\'e}, O., Vicente, S., et al.\ 2017, A\&A, 604, A69\\
Concha-Ram{\'\i}rez, F., Vaher, E., \& Portegies Zwart, S.\ 2019, MNRAS, 482, 732\\
Cuello, N., Dipierro, G., et al.\ 2019, MNRAS, 483, 4114\\
Facchini, S., Clarke, C.~J., \& Bisbas, T.~G.\ 2016, MNRAS, 457, 3593\\
Farias, J.~P., Tan, J.~C., \& Chatterjee, S.\ 2019, MNRAS, 483, 4999\\
Fujii, M.~S., \& Hori, Y.\ 2019, A\&A, 624, A110\\
Hao, W., Kouwenhoven, M.~B.~N., \& Spurzem, R.\ 2013, MNRAS, 433, 867\\
Hands, T.~O., Dehnen, W., et al.\ 2019, MNRAS, 1064\\
Haworth, T.~J., Clarke, C.~J., et al.\ 2018, MNRAS, 481, 452\\
Kounkel, M., Hartmann, L., Loinard, L., et al.\ 2017, ApJ, 834, 142\\
Kuhn, M.~A., Hillenbrand, L.~A., et al.\ 2019, ApJ, 870, 32\\
Lada, C.~J., \& Lada, E.~A.\ 2003, ARAA, 41, 57\\
Malmberg, D., de Angeli, F., Davies, M.~B., et al.\ 2007, MNRAS, 378, 1207\\
Meech, K.~J., Weryk, R., Micheli, M., et al.\ 2017, Nature, 552, 378\\
Olczak, C., Kaczmarek, T., et al.\ 2012, ApJ, 756, 123\\
Olczak, C., Pfalzner, S., \& Spurzem, R.\ 2006, ApJ, 642, 1140\\
Parker, R.~J.\ 2017, Handbook of Supernovae, 2313\\
Pfalzner, S.\ 2009, A\&A, 498, L37\\
Pfalzner, S., Steinhausen, M., \& Menten, K.\ 2014, ApJ, 793, L34\\
Pfalzner, S., \& Bannister, M.~T.\ 2019, ApJ, 874, L34\\
Pfalzner, S., Bhandare, A., et al.\ 2018, ApJ, 863, 45\\
Pfalzner, S., Bhandare, A., \& Vincke, K.\ 2018, A\&A, 610, A33\\
Porras, A., Christopher, M., et al.\ 2003, AJ, 126, 1916\\
Portegies Zwart, S.\ 2019, A\&A, 622, A69\\
Richert, A.~J.~W., Getman, K.~V., et al.\ 2018, MNRAS, 477, 5191\\
Scally, A., \& Clarke, C.\ 2001, MNRAS, 325, 449\\
van Elteren, A., Portegies Zwart, S., et al.\ 2019, arXiv:1902.04652\\
Vincke, K., \& Pfalzner, S.\ 2016, ApJ, 828, 48\\
Vincke, K., \& Pfalzner, S.\ 2018, ApJ, 868, 1\\
Wang, L., Kouwenhoven, M.~B.~N., et al.\ 2015, MNRAS, 449, 3543\\
Wijnen, T.~P.~G., Pols, O.~R., et al.\ 2017, A\&A, 604, A91\\
Winter, A.~J., Clarke, C.~J., et al.\ 2018, MNRAS, 478, 2700\\
Wolff, S.~C., Strom, S.~E., Dror, D., et al.\ 2007, AJ, 133, 1092\\

\normalsize

\end{document}